\documentclass[10pt,letterpaper]{article}
\usepackage{subfig}  
\usepackage{opex3}

\begin{document}

\title{Theory of Slow Light Enhanced Four-Wave Mixing in Photonic Crystal Waveguides}

\author{M. Santagiustina,$^{1}$ C. G. Someda,$^1$, G. Vadal\`a$^1$, \\
S. Combri\'e$^2$, A. De Rossi$^{2}$}
\address{$^1$CNIT, Dipartimento di Ingegneria dell'Informazione, Universit\`a di Padova,
via Gradenigo 6b, Padova, 35131, Italy \\
$^2$Thales Research and Technology, Route D\'epartementale 128, 91767, Palaiseau, France}

\email{marco.santagiustina@unipd.it} 



\begin{abstract}
The equations for Four-Wave-Mixing in a Photonic Crystal waveguide are derived
accurately. The dispersive nature of slow-light enhancement, the impact of
Bloch mode reshaping in the nonlinear overlap integrals and the tensor nature of
the third order polarization are therefore taken into account. Numerical
calculations reveal substantial differences with simpler models, which increase
with decreasing group velocity. We predict that the gain for a 1.3 mm
long, unoptimized GaInP waveguide will exceed 10 dB if the pump power exceeds 1 W.   
\end{abstract}




\section{Introduction}
Slow light (SL) can enable interesting applications in photonics \cite{tuc-khu} 
and microwave-photonics \cite{san09tm} and it is also expected to enhance nonlinear 
phenomena. 
In particular, photonic crystal waveguides (PhCWs) present SL propagation
\cite{baba09natpho} which is predicted to enhance self-phase modulation (SPM)
\cite{bha01pre,sol02pre,kam07jqe}. One intuitive but powerful picture of this
effect represents pulses subject to a spatial compression which locally increases
the power density \cite{kra07jpd}, pretty much as cars in a highway get closer to
each other as their speed is decreased. The dependence of nonlinearity on group
velocity has been recently observed in PhCW for third harmonic generation
\cite{corcoran09} and for SPM and three photon absorption \cite{com09apl,husko09opex}.

Efficient four-wave mixing (FWM) was reported recently in 1.3 mm long, III-V
semiconductor PhCW \cite{eck10ol}, with a conversion efficiency comparable to that
of about 1 m-long highly nonlinear fiber \cite{hase08oc} and about 2 cm-long
chalcogenide fiber photonic wire \cite{pelu08oe}.
That result confirms the theoretical prediction \cite{ebn09oe} of enhanced
(FWM) on the basis of the square group index scaling factor. However, it must be
pointed out that a simple square group index scaling, to model the FWM enhancement,
does not actually take into account several fundamental features of the phenomenon.

The first feature is that the group index in the waveguide is a function of frequency
and, therefore, it is not the same for the various waves involved in the FWM. Observe
in fact in Fig. \ref{sfig:1} the numerically calculated \cite{mpb} group index (red dots)
of a GaInP membrane PhCW. The calculation is carried out with pumps
placed at $N=6$ different wavelengths approaching the band edge:
$\lambda_{1M}=[1570+10\,(M-1)]nm, \, M=1, ..., 6$. 
The second pump wavelength is $\lambda_{2M}= \lambda_{1M} + 2nm$; the signal and
idler frequency are calculated according to 
$\omega_{3M}=\omega_{1M} - \Delta \omega_M$, $\omega_{4M}=\omega_{2M} + \Delta \omega_M$, 
where $\Delta \omega_M = \omega_{1M}-\omega_{2M}$, thus satisfying the FWM frequency
condition $\omega_{1M}+\omega_{2M}=\omega_{3M}+\omega_{4M}$. The waveguide parameters are:
$a=480nm$ (crystal period), $d=0.38a$ (hole diameter) and $h=170\,nm$ (PhC slab height). 

Moreover, the modal superposition of the interacting fields must be also considered.
In PhCWs this overalap is a function of frequency; in particular, as the frequency 
approaches the bandgap, the mode spreads into the hole region as shown
in Fig. \ref{sfig:1}, where the intensity distribution of the electric field within a 
cell of the PhCW is shown at three different wavelengths.
Finally, the PhCWs mode is not constant but periodic along the propagation direction and
so, differently from other photonic waveguides (slabs, fibers etc.), the 
careful determination of the nonlinear effective coefficients is more complicated
\cite{bha01pre,kam07jqe,pan10jqe}.

In this paper, an accurate calculation of all nonlinear effective coefficients necessary 
to evaluate the FWM interaction in the SL regime of a PhCW, is carried out.
The derivation is performed through a perturbation approach, similarly to \cite{mich03pre},
directly from Maxwell's equations.
The resulting SPM coefficient corresponds to that obtained by previous derivations
\cite{bha01pre,kam07jqe}; the XPM coefficient is consistent with the one obtained 
for multimode propagation \cite{pan10jqe}.
The FWM effective coefficients for a PhCW are determined for the first time
to the best of our knowledge.

\section{Derivation of the nonlinear propagation equations}

The starting point of our analysis are Maxwell's equations in the frequency domain
where the linear permittivity  $\varepsilon (\mathbf{r})$ is a spatial function
describing the PhCWs structure and $\mathbf{P}_{NL}(\mathbf{r}, t)$ accounts for
the nonlinear response:
\begin{equation}
\nabla \times \mathbf{E}(\mathbf{r}, \omega) = j \omega \mu \mathbf{H}(\mathbf{r}, \omega), \;\;\;\;
\nabla \times \mathbf{H}(\mathbf{r}, \omega) = -j \omega \varepsilon (\mathbf{r}) 
\mathbf{E}(\mathbf{r}, \omega) -j \omega \mathbf{P}_{NL}(\mathbf{r}, \omega).
\label{eq:4}
\end{equation}
It is assumed that four signals are propagating in the fundamental TE mode of the PhCW,
at frequencies that satisfy the FWM condition $\omega_1+\omega_2=\omega_3+\omega_4$.
The electric and magnetic fields are then expanded as: 
\begin{eqnarray}
\mathbf{E}(\mathbf{r}, \omega) = \frac {1}{2} \sum_{i=-4, \, \neq 0}^4 A_i 
\mathbf{e}(\mathbf{r}, \omega_i) \exp[\imath(\beta_i z)], \;\;\;\; 
\mathbf{H}(\mathbf{r}, \omega) = \frac {1}{2} \sum_{i=-4, \, \neq 0}^4 A_i
\mathbf{h}(\mathbf{r}, \omega_i) \exp[\imath(\beta_i z)], 
\label{eq:2}
\end{eqnarray}
where $A_i$ are the complex amplitudes and the pairs: 
$\mathbf{e}(\mathbf{r}, \omega_i)\exp(\imath\beta_i z) , 
\mathbf{h}(\mathbf{r}, \omega_i)\exp(\imath\beta_i z) $ are the Bloch modes 
at frequencies $\omega_i$ ($\omega_{-i}=-\omega_i$) with 
$\beta_i=\beta(\omega_i)$ 
the propagation constant.
Bloch modes satisfy linear Maxwell's equations individually:
\begin{eqnarray}
\nabla \times \left[ \mathbf{e}_i \exp(\imath\beta_i z) \right] =
\imath \mu\omega_i \mathbf{h}_i \exp(\imath\beta_i z), \;\;\;\;
\nabla \times \left[ \mathbf{h}_i \exp(\imath\beta_i z) \right] =
- \imath \varepsilon  \omega_i \mathbf{e}_i \exp(\imath\beta_i z).
\label{eq:3}
\end{eqnarray}
and also obey the relations:
$ \mathbf{e}_{-i} = \mathbf{e}_{i}^*, \mathbf{h}_{-i} = -\mathbf{h}_i^*$.
For the sake of simplicity, we have omitted the explicit dependence on space and
replaced the frequency $\omega_i$ with the subscript $i$.

We assume that the nonlinearity is small enough so that complex amplitude,
$A_i = A_i (z)$, in the direction $\hat{z}$ is slowly varying in comparison to
$\exp(\imath \beta_i z)$ and to the Bloch mode within the cell
$\{ \mathbf{e}_i, \mathbf{h}_i \}$. 
Furthermore, the Bloch modes are normalized: 
$\int_V (\mathbf{e}_i \times \mathbf{h}^*_i + \mathbf{e}_i^* \times \mathbf{h}_i) \cdot 
\hat{z}\,dV=4a$. Let us stress that, with this choice, $|A_i|^2=P_i$ is the active power
propagating in the $z$ direction at frequency $\omega_i$ \cite{kam07jqe}. 
In the following, we will show that this is the natural choice for normalizing the Bloch
modes when group velocity is substantially different from the phase velocity.
Similarly as in ref. \cite{mich03pre} we consider Eqs. \ref{eq:4} calculated at
frequency $\omega_i$; the second equation is scalarly multiplied by 
$\mathbf{e}_i^* \exp(-\imath\beta_i z)$ and then subtracted from the first, multiplied by
$\mathbf{h}_i^* \exp(-\imath\beta_i z)$. The result is integrated over the volume
of the PhCW unit cell to obtain:
\begin{equation}
\frac{\partial A_i}{\partial z} \int_V \hat{z}\cdot[\mathbf{e}_i \times \mathbf{h}_i^* 
+ \mathbf{e}_i^* \times \mathbf{h}_i]  \, dV = j \omega_i \int_V 
\mathbf{e}_i^* \exp(-\imath\beta_i z) \cdot \mathbf{P}_{NL} (\mathbf{r}, \omega_i)  \, dV
\label{eq:reciproc2}
\end{equation}
Here, we also used the hypothesis that $A_i, \partial A_i / dz$ are slowly varying
functions of $z$ and therefore can be taken constant over one unit cell. 

We now introduce the explicit form of the third-order nonlinear polarization
$\mathbf{P}_{NL} (\mathbf{r}, \omega_i)$. Using the notation of \cite{boyd} it reads:
\begin{eqnarray}
\mathbf{P}_{NL}(\mathbf{r}, \omega_i) = 
\varepsilon_0 \chi^{(3)} (\mathbf{r}; \omega_i; \omega_i, -\omega_i, \omega_i) \vdots \,
\mathbf{E}_i \mathbf{E}_i^* \mathbf{E}_i + \nonumber \\
+ \varepsilon_0 \sum_{j=1, \, \neq i}^4 \left[ 
\chi^{(3)} (\mathbf{r}; \omega_i; \omega_j, -\omega_j, \omega_i)
\vdots \, \mathbf{E}_j \mathbf{E}_j^* \mathbf{E}_i \right] +
\varepsilon_0 \chi^{(3)} (\mathbf{r}; \omega_i; \omega_j, -\omega_l, \omega_k) \vdots
\mathbf{E}_j \mathbf{E}_l^* \mathbf{E}_k,
\label{eq:5}
\end{eqnarray}
where $i,j,k,l=\{1,2,3,4\}$, with the constraint that $\{i,j,k,l\}$ are all 
different in the last term. The susceptibility tensor is real because multi-photon
absorption can be neglected in GaInP at $\lambda \simeq 1.6\mu m$. 
On the right hand side (RHS) of Eq. (\ref{eq:5}), the first term represents SPM, the
summation term XPM, and the last the non-degenerate FWM.
Here, third and other harmonic generations are neglected by assuming that they will
not be phase matched. For the sake of brevity the tensor explicit dependence on
position and frequencies will also be omitted till the end of the derivation.
Inserting Eq. (\ref{eq:5}) in Eq.(\ref{eq:reciproc2}) we obtain: 
\begin{eqnarray}
4a \, \frac{\partial A_i}{\partial z} 
= \frac{\imath \omega_i \varepsilon_0}{4} \left[ |A_i|^2 A_i 
\int_V \mathbf{e}_i^* \cdot \chi^{(3)} \vdots \, \mathbf{e}_i \mathbf{e}_i^* 
\mathbf{e}_i \, dV + \right. \nonumber \\
\left. + \sum_{j=1, \neq i}^4 |A_j|^2 A_i
\int_V \mathbf{e}_i^* \cdot \chi^{(3)} \vdots \, \mathbf{e}_j \mathbf{e}_j^* \mathbf{e}_i 
\, dV + A_j A_l^* A_k \exp(-\imath \sigma_i \Delta \beta z) 
\int_V \mathbf{e}_i^* \cdot \chi^{(3)} \vdots  \, \mathbf{e}_j \mathbf{e}_l^* \mathbf{e}_k 
\, dV \right].
\label{eq:10}
\end{eqnarray}
Here $\Delta \beta = \beta_3 +\beta_4 -\beta_1 -\beta_2$ is the linear phase mismatch
and $\sigma_i=\pm 1$, where the plus (minus) sign applies for $i=3,4$ ($i=1,2$).
In Eq. \ref{eq:10}, the SL enhancement of the nonlinear response is hidden in the
integrals.  In order to make this dependence explicit, we use the identity between the
electromagnetic energy velocity $\mathbf{v}_e$ and the group velocit that holds in 
lossless homogeneous media, in periodic ones \cite{yeh79josa} and in
PhCWs \cite{sakoda_book}. By projecting the energy velocity along
the axis unit vector $\hat{z}$ and using the property that the
space-time average magnetic and electric energies are equal for Bloch modes
\cite{lom05oqe}, $\mu_0/4 \int_V \mathbf{h}_i \cdot \mathbf{h}_i^* = 
1/4 \int_V \mathbf{e}_i \cdot \mathbf{d}_i^*$,
the following is obtained:
\begin{equation}
\mathbf{v}_{ei} \cdot \hat{z} = \frac{1/4 \, \int_V (\mathbf{e}_i \times \mathbf{h}^*_i + 
\mathbf{e}^*_i \times \mathbf{h}_i)\cdot \hat{z}  \,dV}
{1/4 \, \int_V (\varepsilon_0 \varepsilon_r(\mathbf{r}) |\mathbf{e}_i|^2 + 
\mu_0 |\mathbf{h}_i|^2)\,dV}= 
\frac{4a}{2 \int_V \varepsilon_0 \varepsilon_r(\mathbf{r}) |\mathbf{e}_i|^2 \, dV}= 
\frac{2a}{\varepsilon_0 W_i} = v_{gi}. 
\label{eq:15}
\end{equation}
Note that energies appearing above are normalized consistently with the
choice $|A_i|^2=P_i$. We can now normalize the terms on the RHS of Eq. 
(\ref{eq:10}) multiplying them by the factors
$\eta_i^4$ (SPM), $\eta_i^2 \eta_j^2$ (XPM) and $\eta_i \eta_j \eta_k \eta_l$ (FWM),
with $\eta_i=\sqrt{2a/(\epsilon_0\,W_i\,v_{gi})}=1, \, \forall i$. 
The nonlinear coefficients of Eq. \ref{eq:10} are now cast in their canonical
form so that     
the equations governing the FWM in PhCW are obtained
in a form similar to nonlinear fiber optics \cite{agra}:
\begin{equation}
\frac{dA_i}{dz} = \imath \gamma_i |A_i|^2 A_i + 2 \imath \sum_{j=1, \neq i}^4 \gamma_{ij} 
|A_j|^2 A_i + 2 \imath \gamma_{Fi} A_l^* A_j A_k e^{- \imath \sigma_i \Delta \beta z}, 
\;\; i=1,2,3,4,
\label{eq:12}
\end{equation}
with the effective nonlinear coefficients and the relative effective volumes taking
the form:
\begin{eqnarray}
\label{spm}
\gamma_i = \frac{n_2 \omega_i a}{c V_i}; \;\;\;\;
\frac{1}{V_i} = \frac{n_{gi}^2}{W_i^2} 
\int_V \frac{\varepsilon_r}{3 \chi_{xxxx}^{(3)}}
\mathbf{e}_i^* \cdot \chi^{(3)}(\mathbf{r}; \omega_i; \omega_i, -\omega_i, \omega_i) 
\vdots \, 
\mathbf{e}_i \mathbf{e}_i^* \mathbf{e}_i \, dV; \\
\label{xpm}
\gamma_{ij} = \frac{n_2 \omega_i a}{c V_{ij}}; \;\;\;\;
\frac{1}{V_{ij}} = \frac{n_{gi} n_{gj}}{W_i W_j} 
\int_V \frac{\varepsilon_r}{6 \chi_{xxxx}^{(3)}} 
\mathbf{e}_i^* \cdot \chi^{(3)}(\mathbf{r}; \omega_i; \omega_j, -\omega_j,\omega_i)
\vdots \, \mathbf{e}_j \mathbf{e}_j^* 
\mathbf{e}_i \, dV; \\
\label{fwm}
\gamma_{Fi} = \frac{n_2 \omega_i a}{c V_{Fi}}; \;\;\;\;
\frac{1}{V_{Fi}} = \prod_{n=1}^4 \left( \frac{n_{gn}}{W_n} \right)^{1/2}
\int_V \frac{\varepsilon_r}{6 \chi_{xxxx}^{(3)}} 
\mathbf{e}_i^* \cdot \chi^{(3)}(\mathbf{r}; \omega_i; \omega_j, -\omega_l, \omega_k) 
\vdots \, \mathbf{e}_j \mathbf{e}_l^* \mathbf{e}_k \,dV;
\end{eqnarray}
and where $n_2 = 3 \chi_{xxxx}^{(3)}/( 4 \, \varepsilon_r \,\varepsilon_0 \,c)$ 
is the bulk, nonlinear refractive index coefficient for a linear state of
polarization \cite{boyd}. 

So, for each nonlinear effect (SPM, XPM and FWM) we determined:
1) the correct enhancement factor due to SL; 2) the correct overlap 
integral. The obtained SPM and XPM coefficients are consistent to those previosuly 
found \cite{bha01pre,kam07jqe,pan10jqe}; the FWM coefficient is derived for the 
first time to the best of our knowledge.
By observing Eqs. (\ref{spm},\ref{xpm},\ref{fwm}) a general rule can be remarked:
the enhancement factor due to the SL is always given by the geometric mean of the
group indexes of the waves interacting through the tensor $\chi^{(3)}$.

\section{Numerical Results}

\begin{figure}[htbp]
\centering
\subfloat{\label{sfig:1}\includegraphics[width=6.5cm]{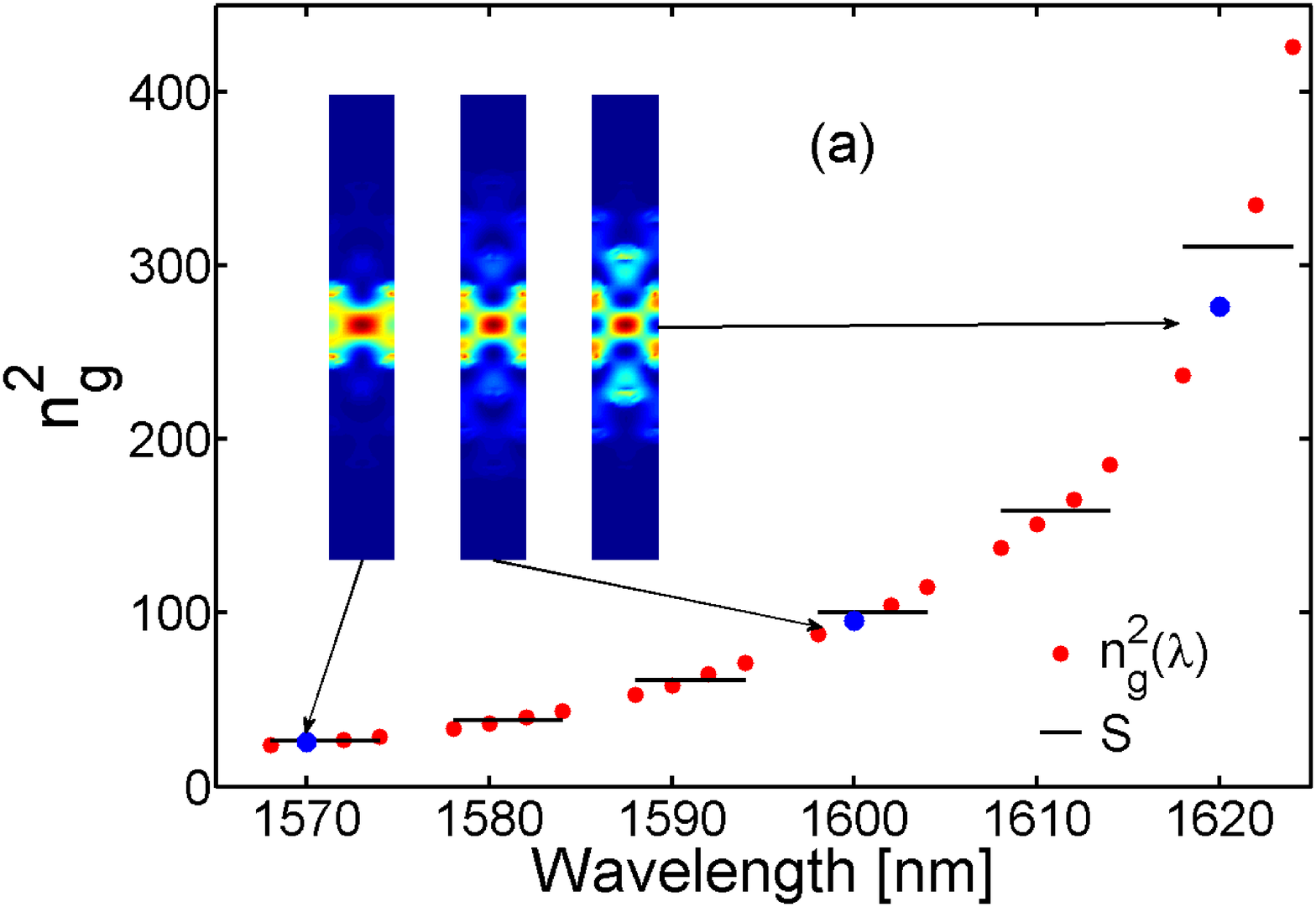}}
\subfloat{\label{sfig:2}\includegraphics[width=6.5cm]{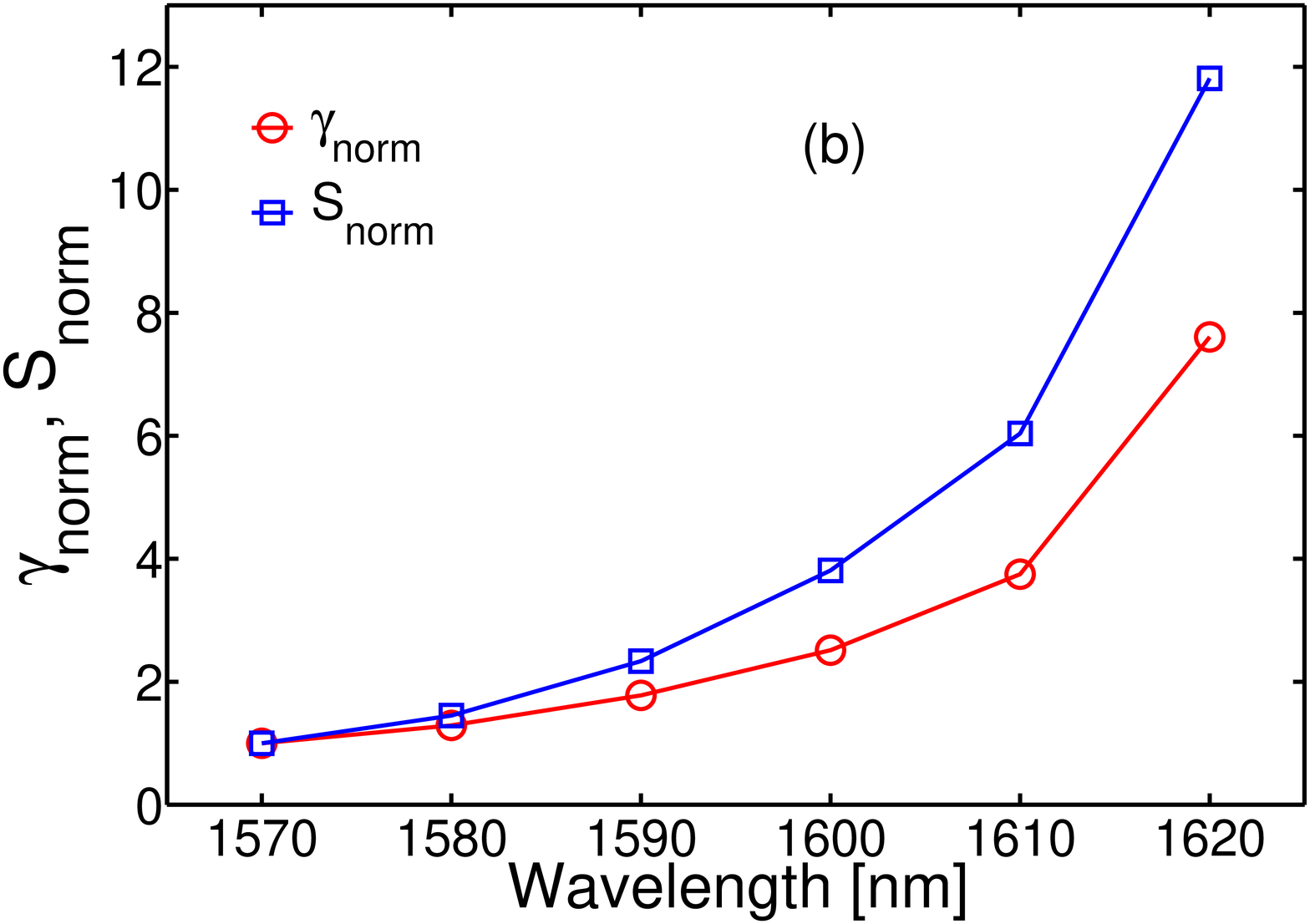}}
\caption{
Comparison with model in ref.\cite{ebn09oe}.\subref{sfig:1} Square of the group index
$n_g^2$ and SL enhancement factor $S$,
as a function of the wavelength. In the insets the intensity of
electric field $\left| \mathbf{e}_i \right|^2$ of the Bloch mode 
within an elementary cell is shown at three different wavelengths.
\subref{sfig:2} 
Wavelength dependence of the SL scaling factor and of the effective FWM
nonlinear coefficient, both normalized to their respective value at $\lambda_{11}=1570\,nm$:
$\gamma_{norm}= \left| \gamma_{F4}(\lambda_{1M})/\gamma_{F4}(\lambda_{11}) \right| $,
$S_{norm}=S(\lambda_{1M})/S(\lambda_{11})$.
}
\end{figure}

The theoretical findings of the previous section are applied to the PhCW of Fig.
\ref{sfig:1}. To the aim of explicitly calculating the tensor
products in Eqs. (\ref{spm},\ref{xpm},\ref{fwm}) the theory of ref. \cite{hut94oc}, that
can be generally applied to zinc-blend semiconductors (group symmetry $\bar{4}3m$),
is exploited and then 
$\chi_{xxxx}^{(3)}= 2 \chi_{xyxy}^{(3)}=2 \chi_{xyyx}^{(3)}= \chi_{xxyy}^{(3)}$.
For signal wavelengths $1.52 \, \mu m < \lambda < 1.62 \, \mu m$ and given that
$E_g \simeq 1.9eV$ for GaInP, the frequencies $\omega_{i,j,k,l}$ at which the 
tensor elements are to be calculated satisfy the condition $0.4 < \hbar \,
\omega_{i,j,k,l} / E_g < 0.43$. Then, although Kleinmann symmetry \cite{boyd} 
is not satisfied, it is found that the dichroism parameter, defined by
$ \chi_{xxyy}^{(3)} / \chi_{xxxx}^{(3)}$, can be approximated to 0.28 \cite{hut94oc}.
Thus all tensors can be determined from the above relations, from the knowledge
of the nonlinear refractive index in GaInP, $n_2 = 10^{-17} m^2/W$ and 
through \cite{boyd}:
\begin{equation}
\mathbf{e}_i^* \cdot \chi^{(3)}(\mathbf{r}) \vdots \, 
\mathbf{e}_j \mathbf{e}_l^* \mathbf{e}_k = \sum_m \left[ e_{im}^* \, D \sum_{nop}
\chi_{mnop}^{(3)}(\mathbf{r}; \omega_i; \omega_j, -\omega_l, \omega_k)  
e_{jn} e_{lo}^* e_{kp} \right] 
\label{tensor}
\end{equation}
where the summations over the indexes $m,n,o,p$ are made on all possible values of the 
coordinate axes $\{x,y,z\}$ and $D$ is the frequency degeneracy factor \cite{boyd}
which represents the number of distinct permutations of the three frequencies
$\{\omega_j, -\omega_l, \omega_k\}$ ($D=3$ for SPM, $D=6$ for XPM and FWM).

We numerically determined, for all previously defined pump and signal wavelengths 
the Bloch mode electrical field and the dispersion relation \cite{mpb}, then calculating 
the FWM SL scaling factor $S=\prod_{k=1}^4 n_{gk}^{1/2}$ and the effective nonlinear 
coefficient $\gamma_{F4}$ according to Eq. (\ref{fwm}).
In Fig. \ref{sfig:1} $S$ can be compared to the square of the group index at the mean
frequency, which is the approximation used in refs. \cite{eck10ol,ebn09oe};
a slight discrepancy appears at the band edge (where the SL effect is strong). 
Note that no particular dispersion engineering of the PhCW has been realized on purpose; 
$\Delta \beta$ could still be reduced by design, to increase FWM efficiency. 
To evaluate the effective nonlinear enhancement of PhCW, in Fig. \ref{sfig:2}
the wavelength dependence of $S$ is compared to that of $\gamma$, revealing that 
the enhancement of FWM coefficient does not follow the pure SL scaling $S$, 
the large difference deriving from the decrease in the modal overlap integrals.

\begin{figure}[htbp]
\centering
\subfloat{\label{sfig:3}\includegraphics[width=6.5cm]{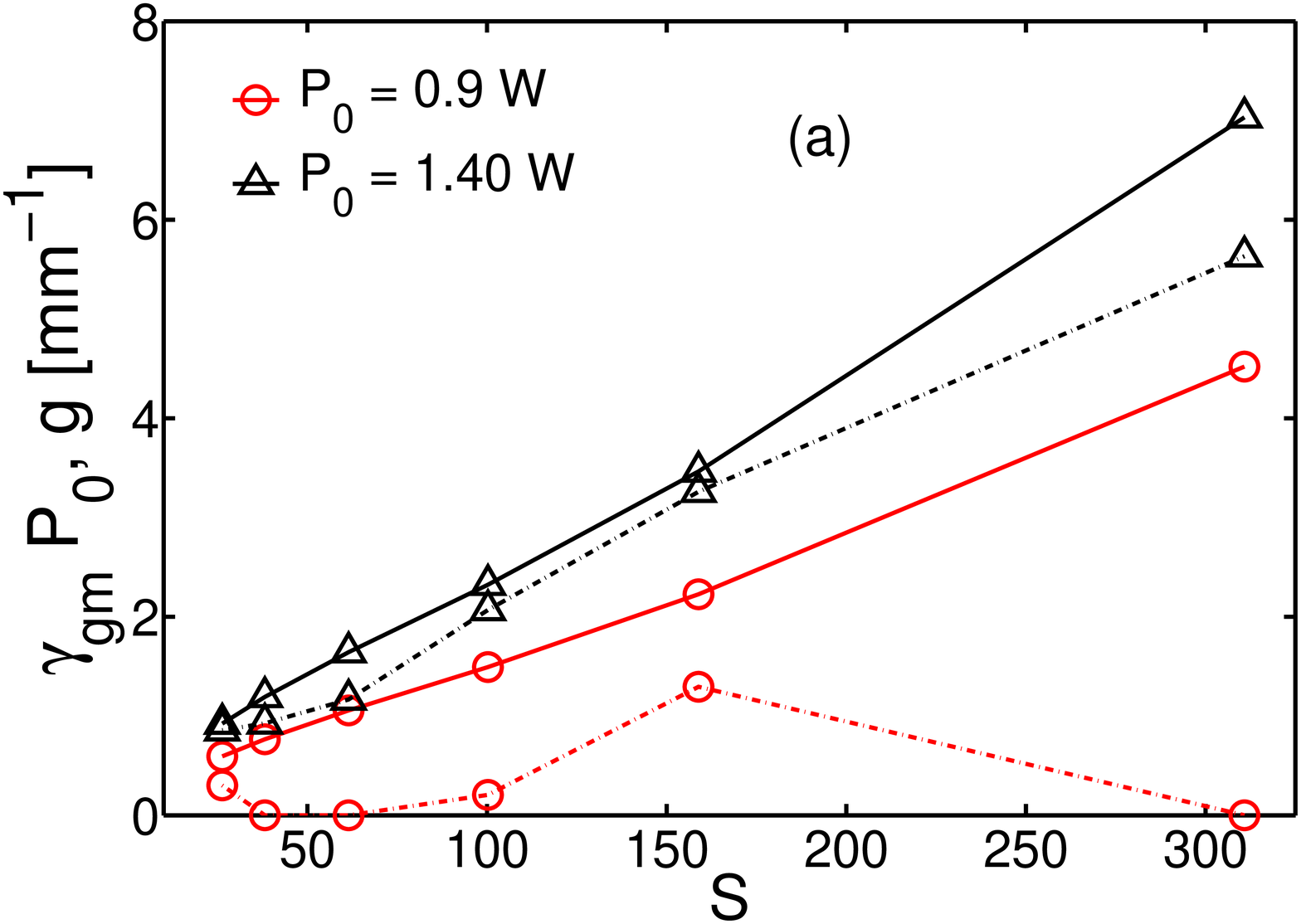}}
\subfloat{\label{sfig:4}\includegraphics[width=6.5cm]{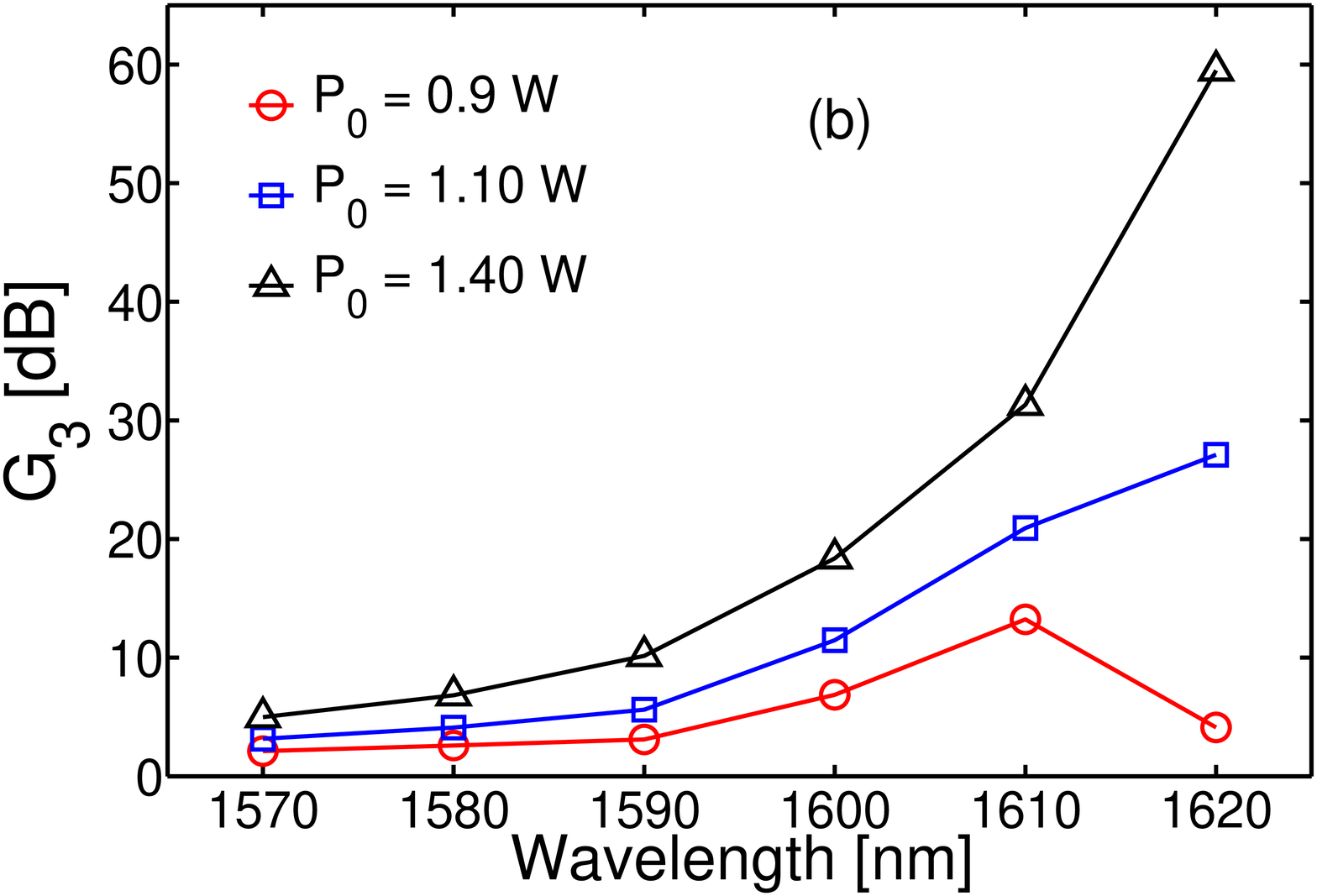}}
\caption{\subref{sfig:3} 
Comparison of the maximum achievable parametric gain coefficient,
$\gamma_{gm} P_0$ (solid curves), and the actually achieved gain, $g$ 
(dashed curves), for two different powers ($P_0=0.9\,W$ circles, $P_0=1.4\,W$ triangles)
as a function of SL scaling factor $S$.
\subref{sfig:4} 
Parametric gain as a function of $\lambda_1$ for a waveguide $L=1.3mm$ long, for
three different pump powers.}
\end{figure}

By following \cite{agra} we can finally determine the nonlinear phase matching and gain
coefficient for optical parametric amplification of the wave at frequency $\omega_3$:
\begin{eqnarray}
\Delta \kappa = \Delta\beta - 
\sum_{i,j=1}^2 [\gamma_i + 2 |i-j| \gamma_{ji} - 2 \gamma_{3i} - 2 \gamma_{4i}]  P_i  = 
\Delta\beta  +\gamma_{pm} P_0, 
\;\;\;\;
g  =\left[ \gamma_{gm}^2 P_0^2 - \frac{\Delta \kappa^2}{4} \right]^{1/2},
\label{etag}
\end{eqnarray}
where the last terms of each of Eqs. (\ref{etag}) are obtained for $P_1=P_2=P_0/2$ and
$\gamma_{gm}^2=\gamma_{F3}\gamma_{F4}^*$. The effective coefficients
$\gamma_{pm}$ and $\gamma_{gm}$ describe the strength  of the nonlinearity contribution,
respectively, to the phase matching and to the maximum gain.
It is remarkable that, differently from fiber optics \cite{agra}, 
$\gamma_{pm} \neq \gamma_{gm}$.
Figure (\ref{sfig:3}) compares the maximum achievable gain coefficient $\gamma_{gm} P_0$
to the actual one $g$, which is limited by the phase mismatch. 
As the pump power increases from $0.9\,W$ to $1.4\,W$ the phase mismatch is almost 
completely canceled by the nonlinear phase terms and the maximum gain is approached. 
This fact leads to a dramatic increase of the FWM gain
$G_3  = P_3(L)/P_3(0)  = 1 + \gamma_{gm}^2 P_0^2/g^2 \sinh^2(gL)$
which is shown in fig. (\ref{sfig:4}) for $L=1.3\;mm$.


\section{Conclusions}
We have derived the nonlinear equations which describe four-wave mixing in photonic crystal
waveguides directly from Maxwell's equations. These equations are exact in the limit 
in which we can neglect the changes that the nonlinearity induces in the Bloch modes 
describing the field in the photonic crystal waveguide, a situation by far verified 
in practice. 
We demonstrate rigorously the explicit dependence of the nonlinear enhancement on the 
group index and that in four-wave mixing (where the fields involved
have different group indexes) the dispersive nature is rigorously accounted for by 
the geometric mean of the group indexes of the modes involved. Moreover, 
we demonstrate a substantial correction arising from Bloch mode reshaping in the 
nonlinear field overlap. Finally we account for the tensor nature of the nonlinear
polarization. As an example, we calculated the gain for a 1.3 mm long waveguide 
operated at moderately small group velocity ($v_g>c/20$). Particularly, when 
$n_g \simeq 12$ and the coupled pump power $> 1 W$, the expected gain exceeds 10 dB
even if the waveguide is unoptimized. 


The research leading to these results has received funding from the European
Community's Seventh Framework Programme (FP7/2007-2011) under grant agreement 
n. 219299 GOSPEL.

\end{document}